# The layer impact of DNA translocation through graphene nanopores

*Wenping Lv*[1], *Maodu Chen*[2], *Ren'an Wu*[1*]

1, CAS Key Lab of Separation Sciences for Analytical Chemistry, National Chromatographic R&A Center, Dalian Institute of Chemical Physics, Chinese Academy of Sciences (CAS), Dalian, 116023, China

2, School of Physics and Optoelectronic Technology, Dalian University of Technology, Dalian 116024, China

**Corresponding Author Footnotes**

Prof. Dr. Ren'an Wu

Tel: +086-411-84379828

Fax: +86-411-84379617

E-mail: wenping@dicp.ac.cn, wurenan@dicp.ac.cn

## Abstract:

Graphene nanopore based sensor devices are exhibiting the great potential for the detection of DNA. To understand the fundamental aspects of DNA translocating through a graphene nanopore, in this work, molecular dynamics (MD) simulations and potential of mean force (PMF) calculations were carried out to investigate the layer impact of small graphene nanopore (2 nm-3 nm) to DNA translocation. It was observed that the ionic conductance was sensitive to graphene layer of open-nanopores, the probability for DNA translocation through graphene nanopore was related with the thickness of graphene nanopores. MD simulations showed that DNA translocation time was most sensitive to the thickness of graphene nanopore for a 2.4 nm aperture, and the observed free energy barrier of PMFs and the profile change revealed the increased retardation of DNA translocation through bilayer graphene nanopore as compared to monolayer graphene nanopore.

## Introduction

Nanopore sequencing has been emerging as a new generation technology for DNA sequencing.[1-5] A variety of biological, solid-stated and biological solid-state hybrid nanopores have been constructed experimentally and/or computationally.[3, 6-13] Graphene is a two-dimensional sheet of carbon atoms arranged in a honeycomb lattice, possessing remarkable mechanical, electrical and thermal properties.[14-16] The subnanometer thickness (0.34 nm) of graphene sheet comparable to the spatial interval of DNA nucleotide suggests a promising DNA sequencing technology with a resolution at single-base level, because of the expected one recognition point in graphene nanopore rather than the multiple contacts in other nanopores.[17] The controllable nanosculpting of graphene nanopore as well as nanobridge and nanogap with few-nanometer precision was first demonstrated by Fischbein and Drndić with a focused electron beam.[18] After that, various graphene nanopores were fabricated with apertures around 5-23 nm (subnanometer thick, monolayer or bilayer graphene),[19] 5-10 nm (1-5 nm thick, *ca.* 3-15 monolayer graphene)[20] and 2-40 nm (0.3-2.7 nm thick, monolayer to eight-layer graphene)[21] etc. Subsequently, the translocation of DNA through nanometer and subnanometer thick graphene nanopores was realized experimentally, with current blockade larger than same size traditional solid-state nanopores.[19-21]

It is interestingly that Golovchenko and co-workers found that the conductance of a monolayer graphene nanopore was proportional to the aperture of nanopore.[19] However, Dekker *et al*. and Schulten *et al*. demonstrated that the pore conductance was proportional to the square of nanopore diameter by means of experiment [21] and MD simulation[22], respectively, similar as the conductance for traditional solid-stat nanopores (e.g. SiN nanopore [23]) which are much thicker. Additionally, it displayed that the conductance of graphene nanopores was impacted by the number of graphene layers, such as for given small graphene nanopores of aperture at 5 nm where ,the conductance of monolayer graphene nanopore was higher than multilayer graphene nanopores,but the conductance of four-layer graphene nanopore was lower than five-layer graphene nanopore.[21] It seemed plausible as terrace effect was observed in multilayer graphene nanopores created by TEM, that the pore edge of multilayer graphene nanopore was much thinner than the layer thickness of the multilayer graphene membrane.[24] On the other hand, Drndic *et al*. used a nanopore constructed in multilayer graphene with pore diameter similar as Golovchenko which gave deeper DNA-induced current blockade than nanopores in monolayer graphene.[20] Though these intriguing differences between monolayer graphene nanopores and multilayer graphene nanopores have been observed, the effect of the thickness of graphene nanopore edge to open-pore conductance and translocation of DNA is still unclear.

By means of the numerical simulation of solving Poisson–Nerst–Planck equation with COMSOL Multiphysics finite element solver, Garaji *et al*. suggested that the graphene monolayer with a 2.4 nm nanopore has the capability to probe DNA molecule (though be modeled as a stiff insulating rod) with a spatial resolution of 0.35 nm (comparable to the distance between two base-pairs in duplex-DNA) at a low translocation speed.[19] In addition, the translocation of DNA through protein and solid-state nanopores have also been successfully investigated by MD simulations at atomic level.[25-28] An all-atom MD simulation was recently carried out to investigate the microscopic kinetics of DNA translocation through graphene nanopore, and the strong effect of external voltage and DNA conformation on ionic current blockade was observed, consistent with experimental observations.[22] Moreover, the discrimination ability of monolayer graphene nanopores to A-T and G-C base pairs (bp) has also been suggested.[22] More recently, researchers found that ionic current blockades produced by different DNA nucleotides were, in general, indicative of the nucleotide type, although very sensitive to the orientation of the nucleotides in the nanopore.[29] These findings allow researcher one step closer to DNA sequencing using graphene nanopores.

Herein, the all-atom MD simulation and potential of mean force (PMF) calculation were carried out to investigate the thickness impact of graphene nanopore toward the pore conductance and DNA translocation. As the terraced effect of multilayer graphene nanopore was observed,[24] in this work, graphene nanopores with subnanometer thickness constructed in monolayer, bilayer and trilayer graphene were applied for MD simulations to investigate the layer impact of graphene nanopores to pore conductance and DNA translocation. We found that 1) the open-pore ionic conductance could be modulated by the layer and aperture of graphene nanopores; 2) DNA translocation could be retard by a thicker and narrower graphene nanopore; 3) the velocity of DNA translocation was sensitive to the layer of graphene nanopores for aperture at 2.4 nm; 4) the free energy barrier (PMF) of DNA fragment through graphene nanopore was increased with the increase of graphene layer.

## Simulation details and Methods:

As the demand of single-base spatial resolution of DNA sequencing by graphene nanopore, [19] graphene nanopores with apertures around 2.4 nm were applied. In total, 9 mono-, bi- and trilayer graphene nanopores with apertures of 2 nm, 2.4 nm and 3 nm were constructed for MD simulation. The schematic diagram of MD simulation model and the modeled structure of graphene nanopores were displayed in Figure 1a and 1b, respectively. The performed MD simulation and corresponding parameters were listed in Table S1 (Supporting Information).

The open-pore *I-V* curves and ionic conductance of these 9 graphene nanopores at different bias voltages (0 V - 3 V) were estimated. The simulation boxes were dimensioned about 6 x 6 x 10 $nm^3$. The pored graphene sheet was positioned in the middle of water boxes in *X-Y* plan. The ionic concentration was 1M of NaCl. The ionic current $I_{(t)}$ was calculated by [22]

$$\mathrm{I}(t) = \frac{\sum_{i=1}^{N} q_i[z_i(t+\Delta t) - z_i(t)]}{\Delta t \mathrm{L}z} \qquad (1)$$

Where, N was the sum runs over all ions, $\Delta t$ was chosen to be 100 ps, and $z_i$ and $q_i$ were the *Z*-coordinate and the charge of ion *i*, respectively.

The ionic conductance was defined as the reciprocal of ionic resistance [22]

$$\mathrm{S} = \frac{\mathrm{I}av}{\mathrm{U}z} \qquad (2)$$

Where, $\mathrm{U}z = \mathrm{L}z\,\mathrm{E}z$ , L$z$ and E$z$ were the length of simulation box and external electric field in *Z*-direction, respectively. I$av$ was the average of ionic current ($I_{(t)}$) during the last 3 ns from a 4 ns MD simulation.

To investigate the thickness effect of graphene nanopores to the translocation of DNA, similar as the simulation for open-pore ionic conductance, the pored graphene sheets are positioned in the middle of three water boxes with dimension of 6 x 6 x 10 $nm^3$. Since a long-chain DNA could enter nanopores in configuration of either unfolded or folded,[19, 30] a short-chain DNA of d-poly$(CAGT)_3$ (12 base-pairs) was employed in MD simulations and manually put in a position where the head base-pair of DNA was just at the entrance of graphene nanopores (see in Figure 1a). Hundreds of $Na^+$ and $Cl^-$ ions were added into the simulation system to make the concentration of NaCl at 1 M and electrically neutral. Four bias voltages of 1, 2, 3 and 4 V in *Z*-direction (perpendicular to graphene membrane, see in Figure 1a) were applied to drive DNA chain through nanopores electrophoretically. In sum, 36 MD simulations with different parameters were set up to study the layer impact of graphene nanopore to DNA translocation (see in Table S1 in Supporting Information).

To understand the inherent difference of DNA translocation in graphene nanopores with different thickness, PMFs of DNA translocation along the centre line of nanopores was calculated further by means of the umbrella sampling combining with the weighted histogram analysis method (WHAM).[31-34] Since the conformational fluctuation of a long chain DNA was so huge during the translocation through a nanopore, it's impossible to get an effective sampling distribution in acceptable simulation time. Thus a shorter DNA fragment composed with only two base-pairs (A*p*T and G*p*C) was built for PMF calculation (see in Figure 1c). Because the orientation DNA in nanopores could impact the interactions between DNA and nanopore, an ideal model that DNA fragment positioned in the center of graphene nanopore was normally employed (Figure 1d).[35] The reaction coordinate was defined as the distance between the centre-of-mass of DNA fragment and a graphene nanopore in *Z*-direction. The length of the calculated reaction coordinate was 1 nm, which could capture both the effects of DNA entrance and translocation in graphene nanopores. In order to ensure the accuracy of PMF calculation, the width of umbrella window was set at 0.1 nm. There were 11 sampling simulations carried out for each PMF calculation, with umbrella potential of

$$w_i(\varphi) = {}^{k}\!/_{2}\,(\varphi - \varphi_i^{c})^2 \qquad (3)$$

Which restrains DNA at the position $\varphi_i^{c}$ ($i$=0, …, 10). A force constant of $k = 1{,}000\ \mathrm{kJ/(mol \cdot nm^2)}$ was selected to ensure the validity of sampling. Three atoms in each nucleobase (diagrammatized in the insets of Figure 1d) were restrained by a two dimension (2D)-harmonic potential with force constant at $1{,}000\ \mathrm{kJ/(mol \cdot nm^2)}$ in *X*- and *Y*-direction.[36, 37] Thus the undesirable conformations such as the nucleobases stack within base-pair or the pairing broken for base-pairs were excluded in sampling simulations. The umbrella positions were recorded every step during simulations, with a total simulation time of 10 ns for each sampling simulation. More than 200 ns sampling simulations were carried out in total. The g_wham program was used to reconstruct the free energy profiles from the umbrella histograms that were collected during umbrella sampling simulations.[31] The umbrella histograms for monolayer and bilayer graphene nanopores were illustrated in Figure S1 and S2, respectively (Supporting Information).

The structures of DNA used in above simulations were all in A-DNA model and built with the program of X3DNA.[38] The graphene nanopores were generated by the VMD program.[39] All MD simulations were carried out by means of GROMACS 4.5 program package.[40] The AMBER99 force field [41] was used to model DNA segments and TIP3P [42] water molecules and ions. The parameters for graphene carbon atoms were those of *sp*2 carbon in benzene in the AMBER99 force field. A harmonic potential with a force constant of $1{,}000\ \mathrm{kJ/(mol \cdot nm^2)}$ was used to constrain the position of carbon atoms near the boundary (see in Figure 3b).[22] The cut-offs of van der Walls (vdW) force were implemented by a switching function starting at a distance of 1.1 nm and reaching to zero at 1.2 nm. The particle mesh Ewald (PME) method was used to calculate the electrostatic interactions with a cut-off distance of 1.4 nm.[43] Three-dimensional periodic boundary conditions (PBC) were applied in simulations. Time step of 2 fs was set. Each simulation included 1,000 steps energy minimization, 50 ps solvent relaxation, 500 ps equilibration with DNA constraint and the production MD with time duration from 4 ns to 10 ns, depending on the requirements in different simulation sections (the detailed parameters for each simulation segment please see Table S1 in Supporting Information).

## Results and discussion:

### ***I-V*** **curves and ionic conductance of open graphene nanopores**

Ionic conductance is an important parameter to describe the migration of ions through a nanopore under an applied electric field.[19-21, 44] The open-pore conductance could directly impact the magnitude of ionic current signal. The effect of ion migration on pore resistance has been observed experimentally.[44] Here the layer effect of graphene nanopores to ionic conductance was investigated.

At first, the current response of open-pore to the bias voltage was studied by monitoring the *I-V* curves. The obtained *I-V* curves were shown in Figure 2a for graphene nanopores with different thicknesses (mono-, bi- and trilayer graphene) and apertures (2 nm, 2.4 nm and 3 nm). On the whole, the current response curves were changed with the increase of graphene layers. While the sensitivities of ionic current to the thickness of graphene nanopores were also impacted by nanopore diameter. For example, the current curves of 2nm nanopores on bilayer and trilayer graphene could not be well distinguished until the applied bias voltages were increased to 3V. While the current curves for wider nanopores (2.4 nm and 3 nm apertures) were obviously sensitive to the add-layers of graphene. As shown in Figure 2a, the noise of ionic current (the error bar) was around 1 nA even no bias voltage applied (caused by self-diffusion of ions and the flexibility of graphene nanopores), and increased with the increase of applied bias voltage (up to 3.5 nA for higher bias voltage). It indicated that the current noise was comparable with the interference of add-layers of graphene and seemed not negligible. However, a recent experiment demonstrated that the electrical noise could be effectively reduced by using a nanopore constructed in a graphene-$Al_2O_3$ nanolaminate membrane.[45]

The ionic conductances (Figure 2b) for the nine graphene nanopores were calculated from the linear fitting of *I-V* curves. The ionic conductances for monolayer graphene nanopores with apertures of 2 nm and 3 nm were about 5 nS and 9 nS, which were consistent with those of experimental and computational measurements.[19, 22] It can be seen that the open-pore ionic conductances with different apertures were decreased when the layer number of graphene nanopore increased from mono-, bi- to trilayer, respectively (Figure 2b). These results indicate that within small aperture and subnanometer thickness scale, the ionic conductance might be modulated by even adding or deducting a layer of carbon atoms on graphene nanopores.

**Layer impact of DNA translocation through graphene nanopores**

To investigate DNA translocation through graphene nanopores with different layers, similar as previous section, nine graphene nanopores on mono-, bi- and trilayer graphene sheets were constructed, with applied bias voltages ranged from 1 V to 4V. As listed in Table S1, 36 simulation systems were established for the investigation of DNA translocation processes through graphene nanopores. Each DNA translocation simulation was repeated for three times by using different initialization seeds to enhance its reliability.

The statistics of DNA translocation events were displayed in Figure 3a. It can be seen that the DNA translocation processes were sensitive to the thickness and aperture of graphene nanopore. As tabled in Figure 3a, symbol " + " denotes the event that DNA could pass through nanopores with remaining as the double-strand structure; symbol " - " denotes the event that DNA could not pass through nanopores or the double-strand structure of DNA was disintegrated (such as unwinding or unzipping [46-48]) during the translocation. As shown in Figure 3a, DNA could pass through a wide and thin nanopore even at the lowest bias voltage. With the increase of the thickness of graphene nanopore, the probability of DNA through nanopore was decreased. Namely, DNA could pass through both 3 nm monolayer and bilayer graphene nanopores under all applied bias voltages (1 V ~ 4 V) in all repeated simulations. While, some “ - ” events occurred in the repeated simulations of DNA translocation in the 3 nm trilayer graphene nanopore. On the other hand, the number of “ + ” events were also decreased with the aperture shrink of nanopores. In all repeat MD simulations, DNA could not pass through the 2 nm nanopore constructed neither on mono-, bi- nor trilayer graphene sheet for bias voltage at 1 V. The possible reason could be the compact interaction between narrow graphene nanopore and DNA molecule, which increased the difficulty for DNA to enter into small nanopores (much detail discussion was given in Section S1 of Supporting Information). These results suggested that the probability of DNA pass through a thick and narrow graphene nanopore should be lower than that DNA pass through a thin and broad graphene nanopore.

It is well known that the translocation of DNA through nanopores could be accelerated by applying a higher bias voltage. As shown in Figure 3a, most of the events for DNA translocation through nanopores abided such rule. Whereas, the number of events for DNA passing through 2.4 nm bilayer and 3 nm trilayer graphene nanopores were not always increased with the enhanced bias voltages, suggesting a complex process for DNA translocating graphene nanopores with other possible unrealized factors. As shown in some observation (Figure S4a in Supporting Information), the head of DNA chain could be transformed to a bended conformation, seemed to be relaxed and a B-type DNA.[49] Subsequently, the bended DNA fallen down and blocked the pore entrance under the drive of applied external electrical field (t=1566 ps, Figure S4a). Thus, the increase of the applied bias voltage would not only accelerate the translocation of DNA as expected, but also sometimes promote the yawing of the deformed DNA in some extend. By the way, a recently published work also found that a higher bias voltage may also lead to the translocation of single-strand DNA slower by trapping the DNA in a conformation unfavorable for translocation.[29] Thus the conformational variations of DNA before entering the pore entrances might be one of the possible reasons to explain above observations (more detailed discussion for DNA translocation failure was presented in Section S2 of Supporting Information).

Herein, the translocation time through nanopores were further analyzed for DNA successfully passing through graphene nanopores (noted as " + " events in Figure 3a). The DNA translocation time of $T_t$ was defined as the duration time for DNA segment (12 bp) to pass through a nanopore from its first base-pair (head) to the last base-pair (end). As shown in Figure 3b, the DNA translocation time was strongly dependent on the applied bias voltage, suggesting an important role of the electrophoretic force for DNA translocation. Comparing DNA translocation in different nanopores, it can be observed that the translocation time of DNA could be modulated by the layer and aperture of graphene nanopores. Where, the translocation time for DNA in 2.4 nm graphene nanopores was most sensitive to the layer of nanopores, with the translocation time of DNA significantly increased with the increased layer of graphene nanopore. While, for the compact nanopores (aperture = 2 nm) and/or the loose nanopores (aperture = 3 nm), the discrimination of DNA translocation time for monolayer and bilayer graphene nanopores were relatively weaker. Moreover, the translocation time for DNA through trilayer graphene nanopores were significantly longer than that through the thinner graphene nanopores for these loose nanopores (aperture > 2 nm). These observations indicated that the thickening of graphene nanopore would reduce the translocation velocity of DNA in nanopores. In addition, a recent published research showed that the rate of "single-base steps"-liked translocation of single-strand DNA seemed increased with the layer numbers of graphene from monolayer to trilayer.[29] These results suggest that modulate the thickness of graphene nanopore might be a potential scheme for DNA sequencing in single-base precision based on graphene nanopores.

**PMFs of DNA translocation through graphene nanopore**

To understand the fundamental mechanism for DNA translocation through graphene nanopores with different thickness, the calculation of PMFs for DNA translocating along the central line of the 2.4 nm monolayer and bilayer graphene nanopores were carried out. The calculation details of PMFs were described in the section of *Simulation details and Methods*. The calculated PMFs and the snapshots of the in-pore states ($A_0$, $B_0$), out-pore states ($A_1$, $B_1$) and the critical intermediate states ($A_b$, $B_t$) were presented in Figure 4. Where, the zero point of free energy were chosen as the reaction coordinate at 1 nm (out-pore states). As shown in Figure 4, the free energy difference of DNA fragment translocation into monolayer graphene nanopore (from state $A_1$ to state $A_0$) was about 5 kJ/mol ($E_1$). While the corresponding average interaction energies (Figure S5a, Supporting Information) between graphene and DNA (GRA-DNA) at state $A_1$ and state $A_0$ were $-2.52 \pm 1.06$ kJ/mol and $-27.08 \pm 6.32$ kJ/mol, respectively. The enhancement of GRA-DNA interaction suggests that the degree of freedom for DNA in nanopores was reduced. In other words, the free energy barrier between states of $A_0$ and $A_1$ was mostly attributed to the entropy decrease due to the constraint of graphene nanopore toward DNA. These results agree with the theories of M. Muthukumar. [50, 51] With the thickening of graphene nanopore, the interactions between graphene and DNA were enhanced (the average GRA-DNA interaction energies for states $B_1$ and $B_0$ were $-6.22 \pm 3.57$ kJ/mol and $-60.59 \pm 6.40$ kJ/mol, respectively, Figure S5b in Supporting Information). Thus a higher free energy barrier about 10 kJ/mol ($E_2$) was observed for the identical DNA in bilayer graphene nanopore with the same aperture. It suggests that the free energy barrier was additionally modulated by interactions between the DNA and the nanopore. [50, 51] These results indicated that the constraints induced free energy barriers in graphene nanopores were increase with the increased of the thickness of graphene nanopores.

Moreover, an obviously free energy trap ($E_{trap}$= -5 kJ/mol) was observed before DNA move into the 2.4 nm bilayer graphene nanopore. As shown in Figure 5, at the deepest point of energy trap, the DNA was just positioned at the entrance of bilayer graphene nanopore (state $B_t$). It indicates that the DNA translocation from state $B_1$ to state $B_t$ was a free energy falling process. The DNA entering the 2.4 nm bilayer graphene nanopore from the pore entrance (from state $B_t$ to state $B_0$) need to overcome a energy barrier about 15 kJ/mol ($|Etrap|+E_2$). It was obviously higher than the free energy difference between state $B_0$ and state $B_1$ ($E_2$= 10 kJ/mol). Interestingly, no obviously energy trap was presented at the entrance of monolayer graphene nanopore. So the free energy barrier of DNA entering monolayer graphene nanopore from pore entrance has no significant difference to that from the state $A_1$ ($E_1$= 5 kJ/mol). Figure 5 also show that a free energy decrease was explicitly existed when DNA shifted from state $A_b$ to state $A_0$. While the free energy profile was monotonically increased for DNA translocation from bilayer graphene nanopore entrance (state $B_t$) into the bilayer graphene nanopore (state $B_0$). These results show that increase the thickness of graphene nanopores could not only enhance the constraint of nanopore to DNA, but also impact the profiles of PMFs of DNA entering the nanopores.

## Conclusions：

A systematic investigation for DNA translocating through graphene nanopores with different layers at subnanometer thickness was carried out by means of molecular dynamics simulation. It was observed by MD simulation that there is the layer impact of graphene nanopores toward DNA translocation and open-pore conductance. Results indicate that within small aperture and subnanometer thickness scale, even add or deduct a layer of carbon atoms on graphene nanopores could impact their open-pore ionic conductance. The study of DNA translocation in graphene nanopores show the probability of DNA pass through a thick and narrow graphene nanopore should lower than that DNA pass through a thin and wide graphene nanopore. Two reasons were proposed for the failed translocation of DNA in nanopores: 1) For the narrow nanopores (aperture = 2 nm), the compact interactions between narrow graphene nanopores and DNA molecules increased the difficulty of DNA entering small nanopores; 2) While for the loose nanopores (aperture > 2 nm), the conformational transform of DNA at pore entrance might induce the failure of DNA translocation. The probabilities of DNA pass through the loose nanopores (aperture > 2 nm) were also decreased with the increase of the thickness of nanopores. In the cases of that DNA successfully passed through graphene nanopores, the velocity of DNA translocation in graphene nanopores seemed to be slowed down by adjusting the thickness and aperture of nanopores. The PMFs analysis showed that the free energy differences for DNA in solution and in nanopores were raised with the increase of layer number of graphene nanopores. The rising of the free energy barriers and the profile change of PMFs could be the fundamental reasons of the increase of the DNA translocation time in bilayer graphene nanopores.

In summary, we would conclude that the adjustment of the thickness of graphene nanopores in subnanometer scale would evidently impact DNA translocation in graphene nanopores. Therefore, the precise control of the layer number of graphene at the edge of nanopores should be a very important aspect for DNA translocation as for the prospect nanopore analysis for DNA.

## Acknowledgement:

This work was supported by the National Natural Science Foundation of China (No. 21175134), the Knowledge Innovation Program of Dalian Institute of Chemical Physics and the Hundred Talent Program of the Chinese Academy of Sciences to Dr. R. Wu.

## Supporting information Available:

All performed MD simulations were listed in Table S1. The umbrella histograms of PMFs calculation for monolayer and bilayer 2.4 nm graphene nanopores were shown in Figure S1 and S2, respectively. In Section S1, the potential reasons of DNA translocation failure in compact graphene nanopores (aperture = 2 nm) were discussed. In Section S2, the potential reasons of DNA translocation failure in loose graphene nanopores (aperture > 2 nm) were discussed. In Figure S3, the snapshots of DNA adhered on the surface of 2 nm monolayer graphene nanopores at 1 V bias voltage were showed (a); the snapshots of the disintegration of the double-strand structure of DNA during it try to pass through the 2 nm monolayer graphene nanopore at 2 V bias voltage were showed (b). In Figure S4, the snapshots of DNA falling down on the surface of 2.4 nm bilayer graphene nanopores at 2 V bias voltage were showed (a); the snapshots of the DNA unwinding/unzipping in 2.4 nm graphene nanopore at 3 V bias voltage were showed (b). In Figure S5, the evolutions of the average interaction energies between DNA and monolayer graphene nanopore (a) and bilayer graphene nanopore (b) long their reaction coordinates were showed.

# Reference:

1. D. W. Deamer and D. Branton, *Accounts Chem Res*, 2002, **35**, 817-825.
2. J. Shendure and H. L. Ji, *Nat Biotechnol*, 2008, **26**, 1135-1145.
3. D. Branton, D. W. Deamer, A. Marziali, H. Bayley, S. A. Benner, T. Butler, M. Di Ventra, S. Garaj, A. Hibbs, X. H. Huang, S. B. Jovanovich, P. S. Krstic, S. Lindsay, X. S. S. Ling, C. H. Mastrangelo, A. Meller, J. S. Oliver, Y. V. Pershin, J. M. Ramsey, R. Riehn, G. V. Soni, V. Tabard-Cossa, M. Wanunu, M. Wiggin and J. A. Schloss, *Nat Biotechnol*, 2008, **26**, 1146-1153.
4. D. Fologea, E. Brandin, J. Uplinger, D. Branton and J. Li, *Electrophoresis*, 2007, **28**, 3186-3192.
5. Y. Lansac, H. Kumar, M. A. Glaser and P. K. Maiti, *Soft Matter*, 2011, **7**, 5898-5907.
6. C. Dekker, *Nat Nanotechnol*, 2007, **2**, 209-215.
7. S. W. Kowalczyk, T. R. Blosser and C. Dekker, *Trends Biotechnol*, 2011, **29**, 607-614.
8. M. Wanunu, S. Bhattacharya, Y. Xie, Y. Tor, A. Aksimentiev and M. Drndic, *ACS nano*, 2011, **5**, 9345-9353.
9. G. Baaken, N. Ankri, A. K. Schuler, J. Ruhe and J. C. Behrends, *ACS nano*, 2011, **5**, 8080-8088.
10. B. M. Venkatesan and R. Bashir, *Nat Nanotechnol*, 2011, **6**, 615-624.
11. A. Meller and D. Branton, *Electrophoresis*, 2002, **23**, 2583-2591.
12. M. Karhanek, J. T. Kemp, N. Pourmand, R. W. Davis and C. D. Webb, *Nano Lett*, 2005, **5**, 403-407.
13. L. J. Steinbock, O. Otto, D. R. Skarstam, S. Jahn, C. Chimerel, J. L. Gornall and U. F. Keyser, *J Phys-Condens Mat*, 2010, **22**.
14. M. J. Allen, V. C. Tung and R. B. Kaner, *Chem Rev*, 2010, **110**, 132-145.
15. A. K. Geim, *Science*, 2009, **324**, 1530-1534.
16. C. N. R. Rao, A. K. Sood, K. S. Subrahmanyam and A. Govindaraj, *Angew Chem Int Edit*, 2009, **48**, 7752-7777.
17. Z. S. Siwy and M. Davenport, *Nat Nanotechnol*, 2010, **5**, 697-698.
18. M. D. Fischbein and M. Drndic, *Appl Phys Lett*, 2008, **93**.
19. S. Garaj, W. Hubbard, A. Reina, J. Kong, D. Branton and J. A. Golovchenko, *Nature*, 2010, **467**, 190-U173.
20. C. A. Merchant, K. Healy, M. Wanunu, V. Ray, N. Peterman, J. Bartel, M. D. Fischbein, K. Venta, Z. T. Luo, A. T. C. Johnson and M. Drndic, *Nano Lett*, 2010, **10**, 2915-2921.
21. G. F. Schneider, S. W. Kowalczyk, V. E. Calado, G. Pandraud, H. W. Zandbergen, L. M. K. Vandersypen and C. Dekker, *Nano Lett*, 2010, **10**, 3163-3167.
22. C. Sathe, X. Q. Zou, J. P. Leburton and K. Schulten, *ACS nano*, 2011, **5**, 8842-8851.
23. M. J. Kim, M. Wanunu, D. C. Bell and A. Meller, *Adv Mater*, 2006, **18**, 3149-+.
24. B. Song, G. F. Schneider, Q. Xu, G. Pandraud, C. Dekker and H. Zandbergen, *Nano Lett*, 2011, **11**, 2247-2250.
25. A. Aksimentiev and K. Schulten, *Biophysical journal*, 2005, **88**, 3745-3761.
26. A. Aksimentiev, J. B. Heng, G. Timp and K. Schulten, *Biophysical journal*, 2004, **87**, 2086-2097.
27. U. Mirsaidov, W. Timp, X. Zou, V. Dimitrov, K. Schulten, A. P. Feinberg and G. Timp, *Biophysical journal*, 2009, **96**, L32-L34.
28. G. W. Slater, C. Holm, M. V. Chubynsky, H. W. de Haan, A. Dube, K. Grass, O. A. Hickey, C. Kingsburry, D. Sean, T. N. Shendruk and L. X. Nhan, *Electrophoresis*, 2009, **30**, 792-818.

29. D. B. Wells, M. Belkin, J. Comer and A. Aksimentiev, *Nano Lett*, 2012, **12**, 4117-4123.
30. J. L. Li, M. Gershow, D. Stein, E. Brandin and J. A. Golovchenko, *Nat Mater*, 2003, **2**, 611-615.
31. J. S. Hub, B. L. de Groot and D. van der Spoel, *Journal of Chemical Theory and Computation*, 2010, **6**, 3713-3720.
32. J. G. Kirkwood, *J Chem Phys*, 1935, **3**, 300-313.
33. G. M. Torrie and J. P. Valleau, *Chem Phys Lett*, 1974, **28**, 578-581.
34. S. Kumar, D. Bouzida, R. H. Swendsen, P. A. Kollman and J. M. Rosenberg, *J Comput Chem*, 1992, **13**, 1011-1021.
35. J. Comer and A. Aksimentiev, *J Phys Chem C*, 2012, **116**, 3376-3393.
36. B. L. de Groot, T. Frigato, V. Helms and H. Grubmuller, *J Mol Biol*, 2003, **333**, 279-293.
37. O. Beckstein and M. S. P. Sansom, *P Natl Acad Sci USA*, 2003, **100**, 7063-7068.
38. X. J. Lu and W. K. Olson, *Nucleic acids research*, 2003, **31**, 5108-5121.
39. W. Humphrey, A. Dalke and K. Schulten, *Journal of molecular graphics*, 1996, **14**, 33-38, 27-38.
40. B. Hess, C. Kutzner, D. van der Spoel and E. Lindahl, *J Chem Theory Comput*, 2008, **4**, 435-447.
41. W. D. Cornell, P. Cieplak, C. I. Bayly, I. R. Gould, K. M. Merz, D. M. Ferguson, D. C. Spellmeyer, T. Fox, J. W. Caldwell and P. A. Kollman, *J Am Chem Soc*, 1995, **117**, 5179-5197.
42. W. L. Jorgensen, J. Chandrasekhar, J. D. Madura, R. W. Impey and M. L. Klein, *J Chem Phys*, 1983, **79**, 926-935.
43. U. Essmann, L. Perera, M. L. Berkowitz, T. Darden, H. Lee and L. G. Pedersen, *J Chem Phys*, 1995, **103**, 8577-8593.
44. C. C. Chen, Y. Zhou and L. A. Baker, *ACS nano*, 2011, **5**, 8404-8411.
45. B. M. Venkatesan, D. Estrada, S. Banerjee, X. Z. Jin, V. E. Dorgan, M. H. Bae, N. R. Aluru, E. Pop and R. Bashir, *ACS nano*, 2012, **6**, 441-450.
46. K. Healy, *Nanomedicine-Uk*, 2007, **2**, 459-481.
47. Q. Zhao, J. Comer, V. Dimitrov, S. Yemenicioglu, A. Aksimentiev and G. Timp, *Nucleic Acids Res*, 2008, **36**, 1532-1541.
48. A. F. Sauer-Budge, J. A. Nyamwanda, D. K. Lubensky and D. Branton, *Phys Rev Lett*, 2003, **90**.
49. X. Zhao and J. K. Johnson, *J Am Chem Soc*, 2007, **129**, 10438-10445.
50. M. Muthukumar, *J Chem Phys*, 1999, **111**, 10371-10374.
51. M. Muthukumar, *Phys Rev Lett*, 2001, **86**, 3188-3191.

## Figures:

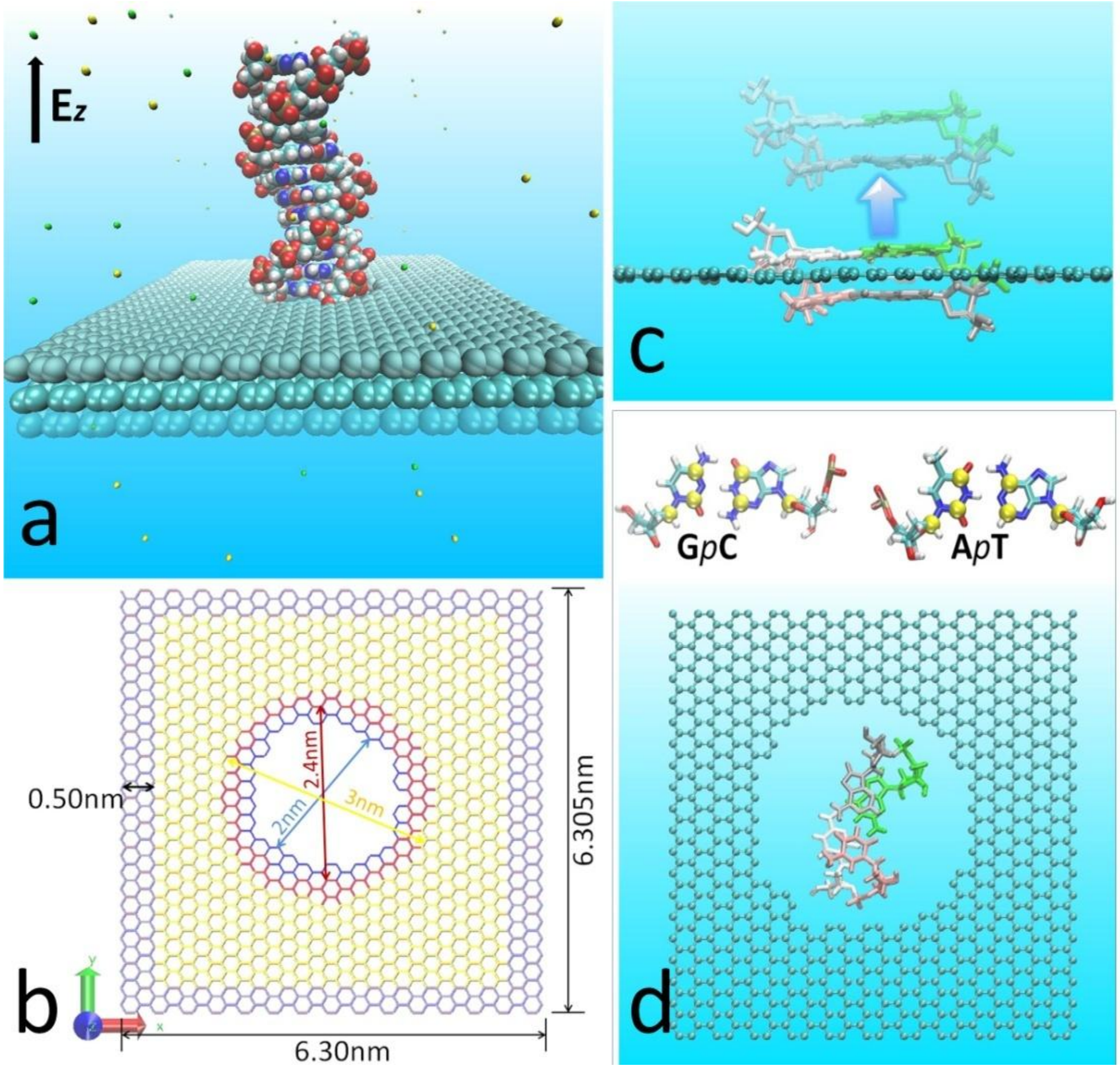

Figure 1. (a) The initial status and scheme of simulation systems for d-poly(CAGT)$_3$ DNA translocation through trilayer graphene nanopores with aperture of 2.4 nm. The ions were plotted as small VDW spheres in colors of yellow ($Na^+$) and blue ($Cl^-$), respectively. (b) The diagrammatizing of nanopores created on monolayer graphene sheet with apertures of 2 nm, 2.4 nm and 3 nm. The blue region represents the restrained atoms. (c) The side view of the in-pore state of DNA and the graphical representation of the reaction coordinate for PMF calculation (the direction of arrow). (d) The plan view of the in-pore state of DNA and the graphical representation of the restrained atoms (shown as yellow balls) in each base-pair.

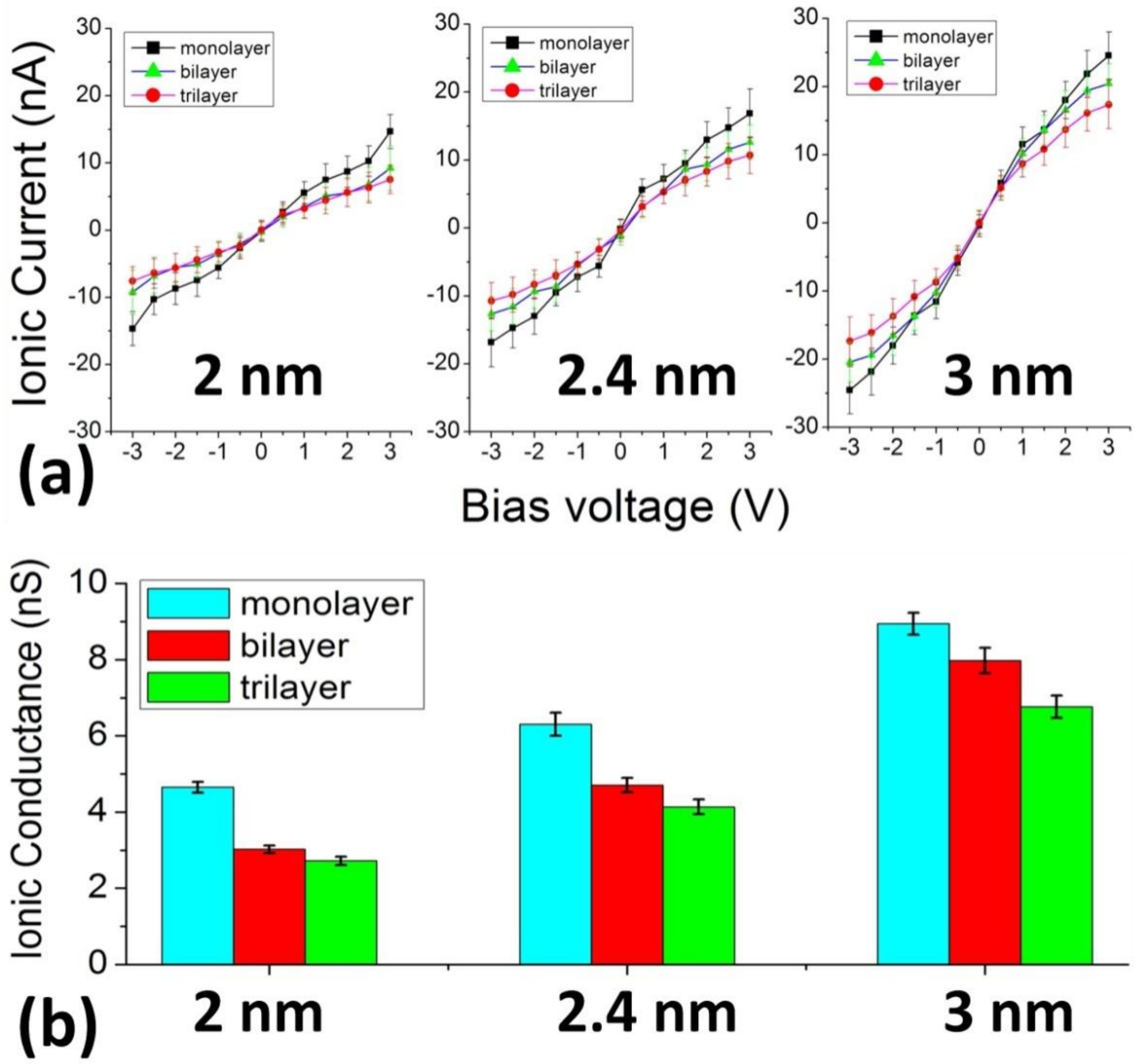

Figure 2. (a) *I*-*V* curves and (b) ionic conductance for graphene nanopores with different apertures (2 nm, 2.4 nm and 3 nm) and thicknesses (mono-, bi- and trilayer graphene). The ionic conductance was obtained by the linear fitting of *I*-*V* curves. The error bars represent the standard deviations.

**Statistics of DNA translocation events**

| | monolayer | | | Bilayer | | | trilayer | | |
|---|---|---|---|---|---|---|---|---|---|
| Aperture (nm) | 2 | 2.4 | 3 | 2 | 2.4 | 3 | 2 | 2.4 | 3 |
| 1 V | − − − | + + − | + + + | − − − | + + − | + + + | − − − | − − − | + + − |
| 2 V | + − − | + + + | + + + | − − − | + − − | + + + | − − − | − − − | + + − |
| 3 V | + + + | + + + | + + + | + + − | + + − | + + + | − − − | + − − | + + + |
| 4 V | + + + | + + + | + + + | + + + | + + + | + + + | − − − | + + + | + + − |

**(a)**

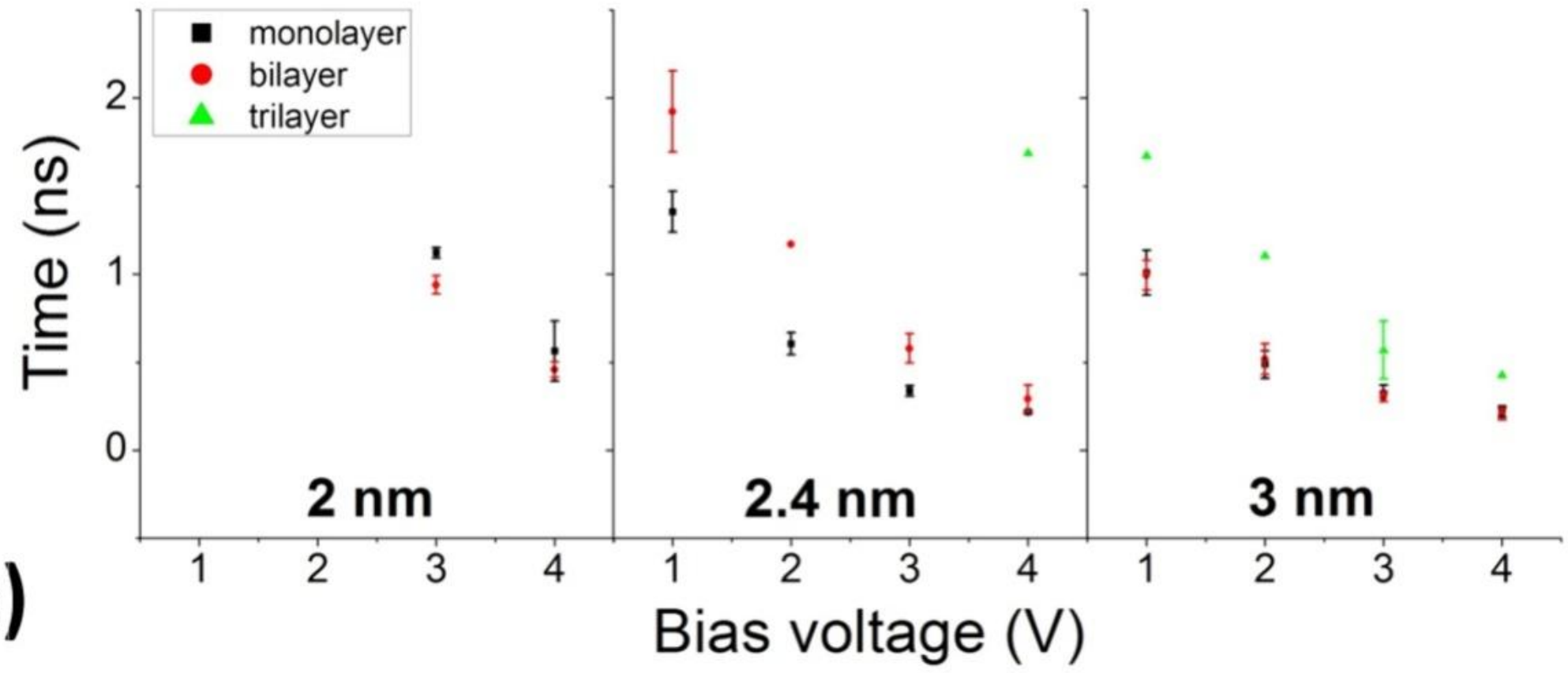

**(b)**

Figure 3. (a) The statistics of DNA translocation events with different nanopores. “ + “ denotes the event that DNA could pass through nanopore; “ - ” denotes the event that DNA could not pass through nanopore. (b) Translocation time of the events (noted as “ + ” in Figure 3a) for DNA successfully passing through graphene nanopores.

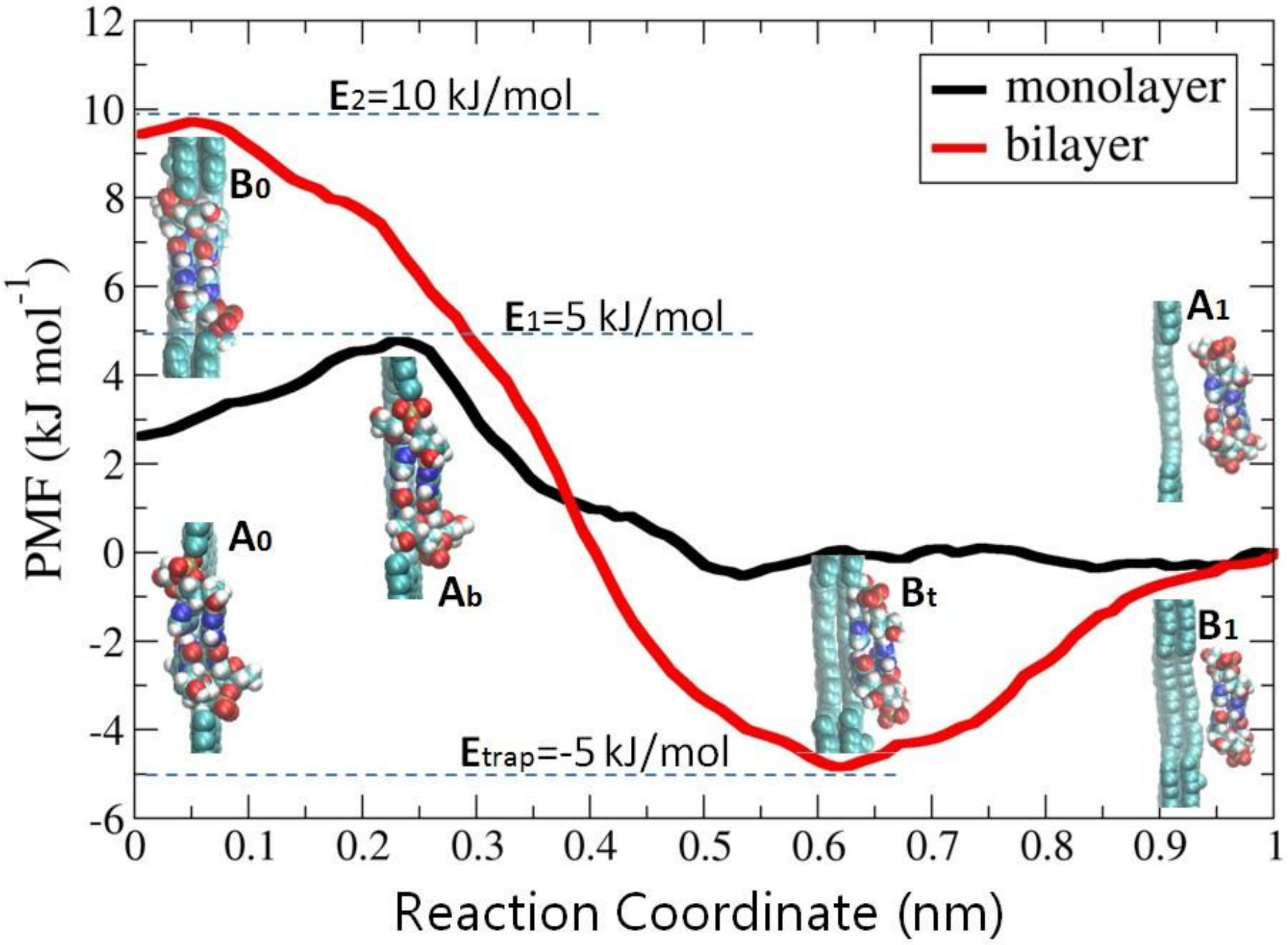

Figure 4. PMFs of DNA translocation in monolayer and bilayer graphene nanopores with aperture of 2.4 nm. The reaction coordinate at 1 nm was chosen as the zero points of free energy. The in-pore ($A_0$, $B_0$), out-pore ($A_1$, $B_1$) and the critical intermediate ($A_b$, $B_t$) states of DNA translocation through monolayer and bilayer graphene nanopores were plotted as insets, respectively.